\documentclass[12pt]{article}
\usepackage{epsfig,graphicx}
\usepackage{fpcp03}

\def\beq{\begin{equation}}
\def\eeq#1{\label{#1}\end{equation}}
\def\eeqn{\end{equation}}

\def\beqa{\begin{eqnarray}}
\def\eeqa#1{\label{#1}\end{eqnarray}}
\def\eeqan{\end{eqnarray}}

\let\bar=\overbar

\def\Dslash{\not{\hbox{\kern-4pt $D$}}}
\def\dslash{\not{\hbox{\kern-2pt $\del$}}}

\def\msb{{\bar{\ssstyle M \kern -1pt S}}}

\def\BB0bar{B^0 {\overline B}^0}
\def\BB0dbar{B_d^0 {\overline B}_d^0}
\def\BB0sbar{B_s^0 {\overline B}_s^0}

\RequirePackage{xspace}

\usepackage{relsize}
\def\babar{\mbox{\slshape B\kern-0.1em{\smaller A}\kern-0.1em
    B\kern-0.1em{\smaller A\kern-0.2em R}}}

\def\Kbar  {\kern 0.2em\overline{\kern -0.2em K}{}\xspace}

\def\Kz    {\ensuremath{K^0}\xspace}
\def\Kzb   {\ensuremath{\Kbar^0}\xspace}
\def\KzKzb {\ensuremath{\Kz \kern -0.16em \Kzb}\xspace}
\def\Kp    {\ensuremath{K^+}\xspace}
\def\Km    {\ensuremath{K^-}\xspace}

\def\KpKm  {\ensuremath{\Kp \kern -0.16em \Km}\xspace}

\def\Dbar    {\kern 0.2em\overline{\kern -0.2em D}{}\xspace}

\def\Dz      {\ensuremath{D^0}\xspace}
\def\Dzb     {\ensuremath{\Dbar^0}\xspace}
\def\DzDzb   {\ensuremath{\Dz {\kern -0.16em \Dzb}}\xspace}
\def\Dp      {\ensuremath{D^+}\xspace}
\def\Dm      {\ensuremath{D^-}\xspace}

\def\DpDm    {\ensuremath{\Dp {\kern -0.16em \Dm}}\xspace}

\def\Bbar    {\kern 0.18em\overline{\kern -0.18em B}{}\xspace}

\def\BB      {\ensuremath{B\Bbar}\xspace} 
\def\Bz      {\ensuremath{B^0}\xspace}
\def\Bzb     {\ensuremath{\Bbar^0}\xspace}
\def\BzBzb   {\ensuremath{\Bz {\kern -0.16em \Bzb}}\xspace}
\def\Bu      {\ensuremath{B^+}\xspace}
\def\Bub     {\ensuremath{B^-}\xspace}

\def\BpBm    {\ensuremath{\Bu {\kern -0.16em \Bub}}\xspace}

\mathchardef\Upsilon="7107
\def\Y#1S{\ensuremath{\Upsilon{(#1S)}}\xspace}% no space before {...}!

\mathchardef\Deltares="7101
\mathchardef\Xi="7104
\mathchardef\Lambda="7103
\mathchardef\Sigma="7106
\mathchardef\Omega="710A

\def\Deltabar{\kern 0.25em\overline{\kern -0.25em \Deltares}{}\xspace}
\def\Lbar{\kern 0.2em\overline{\kern -0.2em\Lambda\kern 0.05em}\kern-0.05em{}\xspace}
\def\Sigbar{\kern 0.2em\overline{\kern -0.2em \Sigma}{}\xspace}
\def\Xibar{\kern 0.2em\overline{\kern -0.2em \Xi}{}\xspace}
\def\Obar{\kern 0.2em\overline{\kern -0.2em \Omega}{}\xspace}
\def\Nbar{\kern 0.2em\overline{\kern -0.2em N}{}\xspace}
\def\Xb{\kern 0.2em\overline{\kern -0.2em X}{}\xspace}

\newcommand{\tev}{\ensuremath{\mathrm{\,Te\kern -0.1em V}}\xspace}
\newcommand{\gev}{\ensuremath{\mathrm{\,Ge\kern -0.1em V}}\xspace}
\newcommand{\mev}{\ensuremath{\mathrm{\,Me\kern -0.1em V}}\xspace}
\newcommand{\kev}{\ensuremath{\mathrm{\,ke\kern -0.1em V}}\xspace}
\newcommand{\ev}{\ensuremath{\mathrm{\,e\kern -0.1em V}}\xspace}
\newcommand{\gevc}{\ensuremath{{\mathrm{\,Ge\kern -0.1em V\!/}c}}\xspace}
\newcommand{\mevc}{\ensuremath{{\mathrm{\,Me\kern -0.1em V\!/}c}}\xspace}
\newcommand{\gevcc}{\ensuremath{{\mathrm{\,Ge\kern -0.1em V\!/}c^2}}\xspace}
\newcommand{\mevcc}{\ensuremath{{\mathrm{\,Me\kern -0.1em V\!/}c^2}}\xspace}
\def\mus  {\ensuremath{\rm \,\mus}\xspace}

\def\mus        {\ensuremath{\,\mu{\rm s}}\xspace}    %% microsecond
\def\pep2{PEP-II}

\def\gsim{{~\raise.15em\hbox{$>$}\kern-.85em
          \lower.35em\hbox{$\sim$}~}\xspace}
\def\lsim{{~\raise.15em\hbox{$<$}\kern-.85em
          \lower.35em\hbox{$\sim$}~}\xspace}

\xspace

\def\jetset74   {\mbox{\tt Jetset \hspace{-0.5em}7.\hspace{-0.2em}4}\xspace}
\begin{document}

%+ \Chapter{}
%+ {Flavour and CP Violation in the Lepton Sector and New Physics}
%+ {St\'ephane~Lavignac}

\begin{flushright}
CERN-TH/2003-266 \\
SACLAY-T03/160
\end{flushright}  
\vskip 1cm 

\Title{Flavour and CP Violation in the Lepton Sector\\
and New Physics \footnote{Invited talk at Flavor Physics and CP Violation
(FPCP 2003), Paris, France, 3-6 June 2003.}}
\bigskip

%+ \addcontentsline{toc}{chapter}{{\it St\'ephane~Lavignac}}
%+ \index{author}{Lavignac, S.} 

%%%%%%%%%%%%%%%%%%%%%%%%%%%%%%%%%%%%%
% Label to flag the first page of your contribution
% Replace Perret by your name starting with a capital letter
%
\label{LavignacStart}

%%%%%%%%%%%%%%%%%%%%%%%%%%%%%%%%%%%%%
% Your name
%
\author{ St\'ephane Lavignac\index{Lavignac, S.}
\footnote{Permanent address:
Service de Physique Th\'eorique, CEA-Saclay, F-91191 Gif-sur-Yvette C\'edex,
France.}
}

%%%%%%%%%%%%%%%%%%%%%%%%%%%%%%%%%%%%%
% Your address
%
\address{Theory Division, CERN\\
 CH-1211 Gen\`eve 23, Switzerland
}

\makeauthor\abstracts{
  We give a pedagogical review of flavour and CP violation in the lepton
sector, with a particular emphasis on new physics -- and in particular
supersymmetric -- contributions to flavour and CP violating observables
involving leptons.
}

\section{Introduction}

In the quark sector, the only source of flavour and CP violation, in the
Standard Model, is the CKM matrix. A number of observables, mainly in the
$K$ and $B$ meson sectors, allow to constrain the mixing angles and the
phase of this matrix and to check the consistency of the CKM picture.
If there is new physics beyond the Standard Model, new sources of flavour
and CP violation are generally present. Their contributions to flavour
and CP violating processes may lead to observable deviations from the
Standard Model predictions. A well-known example of this is the explanation
of the possible discrepancy between $S_{J/\Psi K_S}$ and $S_{\Phi K_S}$
by supersymmetric loop contributions to $B \rightarrow \Phi K_S$
\cite{lavignac-BPhiKs}, while $B \rightarrow J/\Psi K_S$ is dominated by
the Standard Model tree-level contribution.

The situation in the lepton sector is very different. The only experimental
evidence for flavour violation comes from neutrino oscillations. These
imply the existence of a non-trivial lepton mixing matrix, the so-called PMNS
(Pontecorvo-Maki-Nakagawa-Sakata) \cite{lavignac-PMNS} matrix $U$, which is
the analogue of the CKM matrix for leptons.
This non-trivial mixing matrix induces in turn lepton flavour violating (LFV)
processes like $\mu \rightarrow e \gamma$,
$\mu \rightarrow 3\, e$ or $K^0_L \rightarrow \mu e$, and, if it contains
nonzero CP-violating phases, dipole electric moments for charged
leptons. Due to the smallness of the neutrino masses, however,
the corresponding observables are negligibly small
and unaccessible to experiments. For example, the branching
ratio for $\mu \rightarrow e \gamma$ is suppressed by $(m_{\nu_i} / M_W)^4$
\cite{lavignac-mueg}:
\begin{equation}
  \mbox{BR}\, (\mu \rightarrow e \gamma)\ =\ \frac{3 \alpha}{32 \pi}
  \left| \sum_i U^*_{\mu i} U_{ei}\, \frac{m^2_{\nu_i}}{M^2_W} \right|^2\ .
\end{equation}
For $m_{\nu_i} < 1 \mbox{eV}$, one obtains
$\mbox{BR}\, (\mu \rightarrow e \gamma) < 10^{-48}$, well below the present
experimental upper limit.
As for charged lepton electric dipole moments (EDMs), in the absence of
CP-violating phases in the PMNS matrix, they are induced by $\delta_{CKM}$
beyond the 3-loop level \cite{lavignac-de-pospelov91}, which gives
$d_e < 10^{-38}$ e.cm \cite{lavignac-de-hoogeveen90}, well below the
present experimental limit. If the PMNS matrix contains CP-violating phases,
again the EDMs of charged leptons arise at the multiloop level and are
unobservably small.

It follows from these considerations that the observation of any flavour
violating process in the lepton sector other than neutrino oscillations, or
the measurement of charged lepton EDMs, would be a direct signature of new
physics\footnote{New physics could also play a subdominant r\^ole in
neutrino oscillations.}.
This is a strong difference with the quark sector, in which new physics
contributions come in addition to the Standard Model ones.

We have summarized in Table \ref{tab:lavignac-limits} the current upper
limits on some LFV processes and on the
charged lepton EDMs, as well as the expected improvement in the experimental
sensitivity. ``SM prediction'' refers to the prediction of the Standard Model
with Dirac neutrinos. There are many other observables of interest, such as
$K^0_L \rightarrow \mu e$, $K^+ \rightarrow \pi^+ \mu^- e^+$, the rates
of $\mu - e$ conversion in nuclei, or the CP asymmetries in
LFV decays of taus and muons.

%%%%%%%%%%%%%%%%%%%%%%%%%%%%%%%%%%%%%%%%%%%%%%%%%%%%%%

\section{CP violation in neutrino oscillations}

Let us first write the standard parametrization of the PMNS matrix:
\begin{eqnarray}
  U & = & \left( \begin{array}{ccc}
   c_{13} c_{12} & c_{13} s_{12} & s_{13}\, e^{-i \delta} \\
    - c_{23} s_{12} - s_{13} s_{23} c_{12}\, e^{i \delta}
    & c_{23} c_{12} - s_{13} s_{23} s_{12}\, e^{i \delta}
    & c_{13} s_{23} \\
    s_{23} s_{12} - s_{13} c_{23} c_{12}\, e^{i \delta}
    & - s_{23} c_{12} - s_{13} c_{23} s_{12}\, e^{i \delta}
    & c_{13} c_{23}
  \end{array} \right)\ \times \ P\ ,
\end{eqnarray}
where $P = 1$ for Dirac neutrinos, and $P = \mbox{Diag}\,
(1, e^{i \Phi_2}, e^{i (\Phi_3+\delta)})$ for Majorana neutrinos.
This reflects the fact that, in the case of Dirac neutrinos, the PMNS matrix
can be parametrized by 3 angles and  1 CP-violating phase $\delta$, exactly
like the CKM matrix. In the case of Majorana neutrinos, the PMNS matrix
contains 2 additional CP-violating phases $\Phi_2$ and $\Phi_3$
\cite{lavignac-phases-bilenky80,lavignac-schechter80,lavignac-phases-doi81}
which play a r\^ole in neutrinoless double beta decay
\cite{lavignac-phases-doi81}
(for recent discussions, see e.g. Refs. \cite{lavignac-bb0nu}).

\begin{table}[htbp]
\begin{center}
\begin{tabular}{|l|c|l|l|}
\hline 
& SM & & \\
observable & prediction & present experimental limit &
  future expected limit \\
& & & \\
\hline
$\mbox{BR}\, (\mu \rightarrow e \gamma)$ & $< 10^{-48}$ &
  $1.2 \times 10^{-11}$ [MEGA]  & $10^{-14}$ (PSI) \\ 
& & & $10^{-15}$ ($\nu$ factories) \\
$\mbox{BR}\, (\tau \rightarrow \mu \gamma)$ & $< 10^{-48}$
  & $5.0 \times 10^{-7}$ [Belle] & $10^{-8}$ ($B$ factories) \\
$\mbox{BR}\, (\tau \rightarrow e \gamma)$ & $< 10^{-48}$ &
  $2.7 \times 10^{-6}$ [CLEO] & $10^{-8}$ ($B$ factories) \\
$\mbox{BR}\, (\mu \rightarrow e e e)$ & $< 10^{-50}$ &
  $1.0 \times 10^{-12}$ [SINDRUM] & $10^{-16}$ ($\nu$ factories) \\
$\mbox{BR}\, (\tau \rightarrow \mu \mu \mu)$ & $< 10^{-51}$ &
  $8.7 \times 10^{-7}$ [Belle]& ? \\
\hline
$d_e$ (e.cm) & $< 10^{-38}$ & $1.6 \times 10^{-27}$ (Regan 02) &
  $10^{-32}$ (nucl-ex/0109014) \\
$d_\mu$ (e.cm) & $< 10^{-35}$ & $d_\mu = (3.7 \pm 3.4) \times 10^{-18}$
  (Bailey 78) & $10^{-24}$ (BNL) \\
& & & $5 \times 10^{-26}$ ($\nu$ factories) \\
$d_\tau$ (e.cm) & $< 10^{-34}$ & $-2.2 < \mbox{Re}(d_\tau) < 4.5\, (10^{-17})$
  & \\
& & $-2.5 < \mbox{Im}(d_\tau) < 0.8\, (10^{-17})$ [Belle] & ? \\
\hline
\end{tabular}
\caption{Current experimental limits on some flavour and CP violating
observables in the lepton sector.}
\label{tab:lavignac-limits}
\end{center}
\end{table}

The only source of information we have so far on the PMNS matrix are neutrino
oscillation experiments, which also constrain the squared mass differences
$\Delta m^2_{ij} \equiv m^2_{\nu_i} - m^2_{\nu_j}$. $\theta_{23}$ and
$\Delta m^2_{32}$ (resp. $\theta_{12}$ and $\Delta m^2_{21}$) are
associated with oscillations of atmospheric (resp. solar) neutrinos;
both have been found to be large, and possibly maximal for $\theta_{23}$.
The third angle $\theta_{13}$ has not been measured yet, but is constrained
to be smaller than the Cabibbo angle by nuclear reactor experiments
\cite{lavignac-chooz}.
A recent 3-neutrino fit \cite{lavignac-gonzalez03} of all available oscilation
data \cite{lavignac-cavata-fpcp03} gives the following allowed ranges of
parameters at the $1 \sigma$ ($3 \sigma$) confidence level\footnote{Since then,
the SNO collaboration has published new neutral current data which strongly
disfavour the high $\Delta m^2_{21}$ region \cite{lavignac-SNOsalt}.}:
\begin{eqnarray}
  & (1.5)\, 2.2 < \Delta m^2_{32} / 10^{-3}\, \mbox{eV}^2 < 3.0\, (3.3)\ , & \\
  & (0.45)\, 0.75 < \tan^2 \theta_{23} < 1.3\, (2.3)\ , &
\end{eqnarray}
\begin{eqnarray}
  & (5.4)\, 6.7 < \Delta m^2_{21} / 10^{-5}\, \mbox{eV}^2 < 7.7 (10)
  \quad \mbox{and} \quad
  (14) < \Delta m^2_{21} / 10^{-5}\, \mbox{eV}^2 < (19)\ , & \\
  & (0.29)\, 0.39 < \tan^2 \theta_{12} < 0.51\, (0.82)\ , &
  \label{eq:lavignac-theta_12}
\end{eqnarray}
\begin{equation}
   \sin^2 \theta_{13} < 0.02\, (0.052)\ .
\end{equation}
The mixing pattern in the lepton sector is very different from the quark
sector, which has only small mixing angles, but the mass hierarchy is much
less pronounced in the neutrino sector than in the up and down quark sectors.

CP is violated in neutrino oscillations if oscillation probabilities are
different for neutrinos and antineutrinos of the same flavour, {\it i.e.}
$P_{\nu_\alpha \rightarrow \nu_\beta} \neq
P_{\bar \nu_\alpha \rightarrow \bar \nu_\beta}$ for any two lepton flavours
$\alpha$ and $\beta$ \cite{lavignac-cabibbo78}. The oscillation probabilities
in vacuum depend on the entries of the PMNS matrix $U_{\alpha i}$, on the mass
squared differences $\Delta m^2_{ij}$ and on the ratio $L/E$,
where $E$ is the neutrino energy and $L$ the distance travelled by the
neutrino between the production and detection points. For any two distinct
neutrino flavours $\alpha$ and $\beta$:
\begin{eqnarray}
  P_{\nu_\alpha \rightarrow \nu_\beta
  (\bar \nu_\alpha \rightarrow \bar \nu_\beta)} &
  = & 
  -\, 4\, \sum_{i<j} \mbox{Re}
  \left( U_{\alpha i} U^\star_{\beta i} U^\star_{\alpha j} U_{\beta j} \right)
  \sin^2 \left( \frac{\Delta m^2_{ji} L}{4 E} \right)  \nonumber \\
  & & \pm\ 2 J \left[\, \sin \left( \frac{\Delta m^2_{21} L}{4 E} \right)
  +\ \sin \left( \frac{\Delta m^2_{32} L}{4 E} \right)
  -\ \sin \left( \frac{\Delta m^2_{31} L}{4 E} \right) \right]\ ,
\end{eqnarray}
where the last term is CP-odd and takes a minus sign for
$\nu_\alpha \rightarrow \nu_\beta$ and $(\alpha,\beta,\gamma)$ (with
$\gamma \neq \alpha, \beta$) an even permutation of $(e, \mu, \tau)$,
and a plus sign for $\bar \nu_\alpha \rightarrow \bar \nu_\beta$ and
$(\alpha,\beta,\gamma)$ an even permutation of $(e, \mu, \tau)$. $J$
is the Jarlskog invariant \cite{lavignac-jarlskog85} associated with
the PMNS matrix:
\begin{equation}
  J\ \equiv\ \mbox{Im} \left( U_{e2} U^\star_{\mu 2} U^\star_{e 3} U_{\mu 3}
  \right)\ =\ \frac{1}{8} \cos \theta_{13}\, \sin 2 \theta_{23}\,
  \sin 2 \theta_{13}\, \sin 2 \theta_{12}\, \sin \delta\ .
\end{equation}
We immediately see that the Majorana phases $\Phi_2$ and $\Phi_3$
drop from the oscillation formulae, so that CP violation in oscillations
probes only the ''CKM-like'' phase $\delta$
\cite{lavignac-phases-bilenky80,lavignac-phases-doi81}.
Using the hierarchy $\Delta m^2_{21} \ll \Delta m^2_{32}$, one can write:
\begin{equation}
  P_{\nu_\alpha \rightarrow \nu_\beta} -
  P_{\bar \nu_\alpha \rightarrow \bar \nu_\beta}\ \simeq\
  \pm\ 8 J \left( \frac{\Delta m^2_{21} L}{2 E} \right)
  \sin^2 \left( \frac{\Delta m^2_{31} L}{4 E} \right)\ ,
\end{equation}
with a minus sign for the ``golden channel'' $\nu_e \rightarrow \nu_\mu$.
Like in the quark sector, CP violation effects are proportional to the
Jarlskog invariant $J$, which is itself proportional to $\sin \delta$ and
$\sin \theta_{13}$. We also see that the CP asymmetry in neutrino oscillations
is proportional to $\Delta m^2_{21} L / 2 E$.
The conditions for observing CP violation
effects in oscillations are therefore the following: (i) $\Delta m^2_{21}$
and $\theta_{12}$ should be large, which we know is the case since the
KamLAND experiment has identified the so-called LMA (Large Mixing Angle)
solution, characterized by both a large $\theta_{12}$ and a high
$\Delta m^2_{21}$, as the origin of the solar neutrino deficit observed
on the Earth; (ii) $\theta_{13}$ should not be too small; (iv) the
CP-violating phase $\delta$ should be large; and (v) the baseline should be
long enough so that subdominant oscillations, which are governed by the
solar squared mass difference, can develop.

The best experimental conditions for observing CP violation in oscillations
would be provided by a neutrino beam produced from muon decays at a neutrino
factory, and a very long baseline (typically $3000$ km or $7000$ km). For such
large distances, matter effects \cite{lavignac-MSW} -- which induce an apparent
CP asymmetry, due to the fact that matter effects are different for neutrinos
and antineutrinos -- must be taken into account. Another, less ambitious
option is to use neutrino superbeams (which could be produced by JHF in Japan,
NuMI in Fermilab or the SPL at CERN). These beams are characterized by a
lower energy and therefore allow for CP violation searches on shorter
baselines (resp. $130$, $300$ and $730$ km). The SPL superbeam could be
used in combination with a ``beta beam''. For a review on these projects,
see e.g. Ref. \cite{lavignac-nufact}.

%%%%%%%%%%%%%%%%%%%%%%%%%%%%%%%%%%%%%%%%%%%%%%%%%%%%%%%%%%%%%%%%%%%%%%%%

\section{Lepton flavour violating processes}
%-------------------------------------------

As already mentioned in the introduction, LFV processes are unobservable
in the Standard Model with Dirac neutrinos, and more generally if the only
source of flavour violation at low energy is the PMNS matrix.
On the other hand, most extensions of the Standard Model include
new sources of flavour violation (and of CP violation) which, depending
on the case, may be related to the mechanism responsible for neutrino
masses or not. Let us mention, as examples of such extensions, models
in which Majorana neutrino masses are generated radiatively or through
couplings to an $SU(2)_L$ Higgs triplet, supersymmetric extensions of
the Standard Model with or without a seesaw mechanism, and supersymmetric
Grand Unified Theories (GUTs). In the following, we discuss lepton flavour
violation in models with radiative generation of neutrino masses and in
supersymmetric extensions of the Standard Model.

\subsection{Models with radiative generation of neutrino masses}
%...............................................................

In this subsection, we consider models in which neutrino masses are
generated at the quantum level by non-standard interactions of the
neutrinos. These interactions necessarily violate lepton flavour and
therefore induce LFV processes, generally at much larger a rate than
the PMNS matrix itself, although they may still be too weak to be
observable.

The prototype of models with radiative generation of neutrino masses,
the Zee model \cite{lavignac-zee80}, is now excluded by solar neutrino
experiments. We could consider more sophisticated models that pass all
experimental data; however this would only complicate our discussion of
LFV processes, and we prefer to stick to the Zee model. 
This model has two identical Higgs doublets $H$, $H'$ and a charged Higgs
$SU(2)_L$ singlet $h^+$, which couples to two lepton doublets with
antisymmetrized $SU(2)_L$ indices:
\begin{equation}
  f_{\alpha \beta} \left( \nu^T_{L \alpha} C e_{L \beta}\,
  -\, e^T_{L \alpha} C \nu_{L \beta} \right) h^+\ +\ \mbox{h.c.}\ ,
\end{equation}
where $C$ is the charge conjugation matrix, and the couplings
$f_{\alpha \beta} = - f_{\beta \alpha}$ are antisymmetric,
hence flavour violating. Neutrinos are strictly massless at
the tree level, but (Majorana) neutrino masses are induced by the
$f_{\alpha \beta}$ couplings at the one-loop level. The one-loop
neutrino mass matrix has the following structure:
\begin{equation}
  M_\nu\ =\ m \left( \begin{array}{ccc}
    0 & a & b \\
    a & 0 & c \\
    b & c & 0
  \end{array} \right)\ ,
\end{equation}
where $a = f_{e \mu} (\frac{m_\mu}{m_\tau})^2$, $b = f_{e \tau}$
and $c = f_{\mu \tau}$, and the mass scale $m$ depends on the physical
charged Higgs boson masses and mixing angles. This structure implies
a very large solar mixing angle,
$\tan^2 \theta_{12} = 1 - \frac{\Delta m^2_{21}}{2 \Delta m^2_{32}}
\approx 0.98$, which is now excluded by solar neutrino data
(see Eq. (\ref{eq:lavignac-theta_12})).
As mentioned above, the $f_{\alpha \beta}$ couplings induce LFV processes
at the one-loop level; in particular the branching ratio for
$\mu \rightarrow e \gamma$ is found to be \cite{lavignac-petcov82}:
\begin{equation}
  \mbox{BR}\, (\mu \rightarrow e \gamma)\ =\ \frac{\alpha}{48 \pi}\
  \left( \frac{f_{e \tau} f_{\mu \tau}}{\bar M^2 G_F} \right)^2\ ,
\end{equation}
where $\bar M$ is a function of the physical charged Higgs boson masses,
and $G_F$ is the Fermi constant. For values of the parameters relevant
for neutrino masses, e.g. $f_{\alpha \beta} \sim 10^{-5}$ and $M \sim 1$ TeV,
one typically finds $\mbox{BR}\, (\mu \rightarrow e \gamma) \sim (10^{-25}
- 10^{-30})$. This is well above the contribution of the PMNS matrix itself,
although still orders of magnitude below the experimental limit. The reason
is that small neutrino masses require small values of $f_{\alpha \beta} /
\bar M$. A similar conclusion holds for models in which neutrino Majorana
masses are generated from Yukawa couplings involving an $SU(2)_L$ Higgs
triplet \cite{lavignac-schechter80,lavignac-gelmini81}; however, one can
play with the value of the triplet vev so as to increase the value of the
flavour violating Yukawa couplings.

There are many other models which realize the idea of generating neutrino
masses through loop diagrams, and may lead to larger branching ratios for
LFV processes. Let us just mention an interesting possibility that arises
within supersymmetric extensions of the Standard Model without $R$-parity
\cite{lavignac-Rp}, a discrete symmetry usually imposed in order to forbid
dangerous baryon number and lepton number violating couplings. In these
theories, lepton number violating couplings\footnote{Since the simultaneous
presence of baryon number and lepton number violating couplings could lead
to an unacceptably short proton lifetime, we assume baryon number
conservation.} such as $\lambda_{ijk} \tilde e_{jL} \bar e_{kR} \nu_{iL}$
and $\lambda'_{ijk} \tilde d_{jL} \bar d_{kR} \nu_{iL}$, where $\tilde e_L$
and $\tilde d_L$ are the supersymmetric scalar partners of the left-handed
charged leptons and down quarks, respectively, induce (Majorana) neutrino
masses at the one-loop level \cite{lavignac-hall84} and can lead to a viable
mass and mixing pattern. They also contribute to a number of LFV processes
\cite{lavignac-hall84} which could well be accessible experimentally, such as
$\mu \rightarrow e \gamma$, $\mu \rightarrow e e e$ and $\mu - e$
conversion in nuclei (see e.g. Ref. \cite{lavignac-gouvea01}).
The phenomenology of supersymmetry without $R$-parity
is actually very rich and signatures at high-energy colliders are also expected
if $R$-parity violation is responsible for neutrino masses.

\subsection{Supersymmetric extensions of the Standard Model}
%...........................................................

In supersymmetric extensions of the Standard Model like the MSSM (Minimal
Supersymmetric Standard Model), new sources of lepton flavour violation,
and of CP violation, can be present in the slepton (the supersymmetric scalar
partners of the leptons) sector. Indeed, while at the
supersymmetric level sleptons and leptons are degenerate in mass,
supersymmetry breaking generates new contributions to the slepton mass
matrices of the form:
\begin {equation}
  (m^2_{\tilde L})_{\alpha \beta}\, \tilde L^\dagger_\alpha \tilde L_\beta\
  +\ (m^2_{\tilde e})_{\alpha \beta}\, \tilde e^\dagger_{R \alpha}
  \tilde e_{R \beta}\ + \left( A^e_{\alpha \beta} v_d\,
  \tilde e^\dagger_{R \alpha} \tilde e_{L \beta}\  +\ \mbox{h.c.} \right)\ ,
\label{eq:lavignac-soft}
\end{equation}
where $\tilde L^T_\alpha = \left( \tilde \nu_{L \alpha}\,
\tilde e_{L \alpha} \right)$ (resp. $\tilde e_{R \alpha}$) is the
supersymmetric partner of the lepton doublet
$L^T_\alpha = \left( \nu_{L \alpha}\,e_{L \alpha} \right)$ (resp. of the
lepton singlet $e_{R \alpha}$), $m^2_{\tilde L}$ and $m^2_{\tilde e}$
are $3 \times 3$ hermitean mass matrices, $A^e_{\alpha \beta}$ is a
$3 \times 3$ complex matrix, and $v_d$ is the vev of $H_d$, the Higgs doublet
which couples to down quarks and charged leptons in the MSSM. The last term
in Eq. (\ref{eq:lavignac-soft}) mixes ''left'' and ''right'' states and is
known as A-term. Since the mechanism responsible for supersymmetry breaking
is not known, the soft supersymmetry breaking parameters $m^2_{\tilde L}$,
$m^2_{\tilde e}$ and $A^e_{\alpha \beta}$ are arbitrary matrices in flavour
space. In particular, they need not be diagonal in the flavour basis defined
by the charged lepton mass eigenstates, and this results in flavour
violating couplings of sleptons to leptons and charginos/neutralinos (mass
eigenstate combinations of the supersymmetric partners of the gauge and
Higgs bosons), such as $\tilde e^+ \mu^- \tilde \chi^0$ or
$\tilde{\bar \nu}_e\, \mu^- \tilde \chi^+$. These couplings induce large
contributions to LFV processes \cite{lavignac-FCNC-susy}, potentially
above the present experimental limit.

%%%%%%%%%%%%%%%%%%%%%%%%%%%%%%%%%%%%%%%%%%%%%%%%%%%%%%%%%%%%%%%%%%%%%%%%%%%
\begin{figure}[htb]
\begin{center}
\epsfig{file=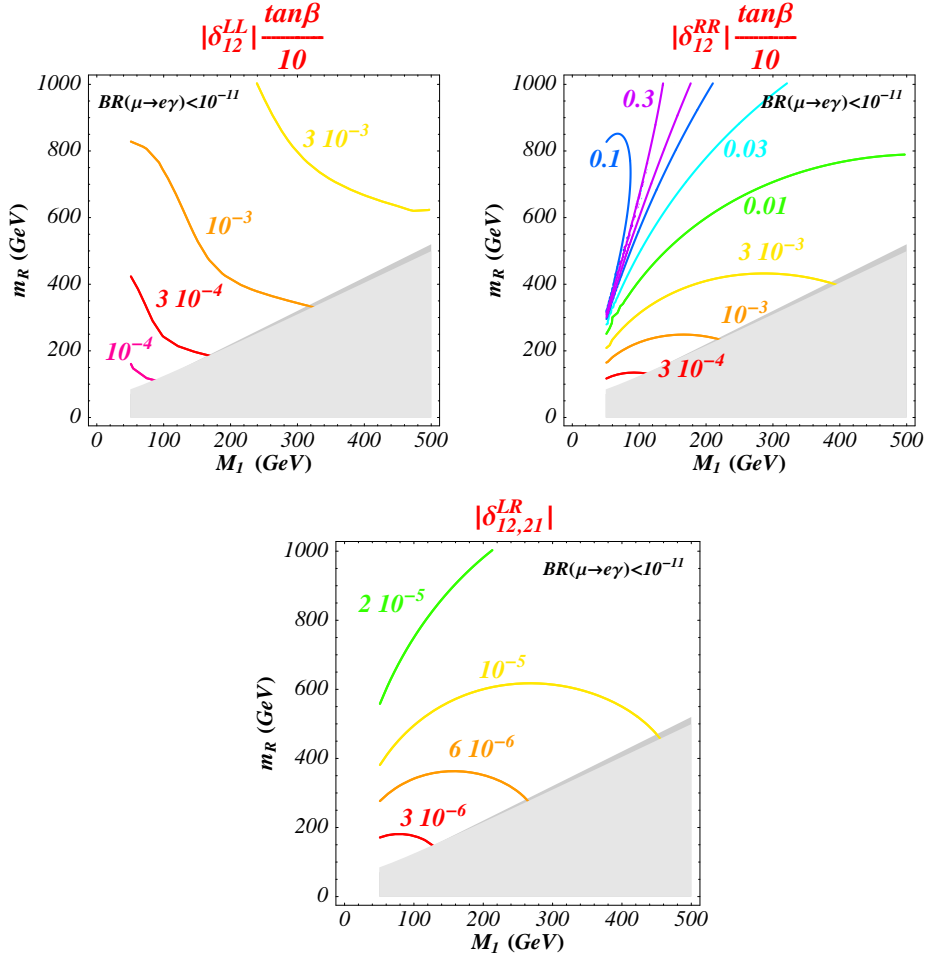,height=130mm}
\caption{Upper bounds on $|\delta^{LL}_{12}| \frac{\tan \beta}{10}$,
$|\delta^{RR}_{12}| \frac{\tan \beta}{10}$ and
$|\delta^{LR}_{12,21}|$ as functions of the average right slepton mass
$m_R$ and of the bino (supersymmetric partner of the hypercharge gauge boson)
mass $M_1$. The light (dark) grey region is unphysical (excluded by cosmology).
From Ref. \cite{lavignac-savoy02}.}
\label{fig:lavignac-delta12}
\end{center}
\end{figure}
%%%%%%%%%%%%%%%%%%%%%%%%%%%%%%%%%%%%%%%%%%%%%%%%%%%%%%%%%%%%%%%%%%%%%%%%%%%

In practice, one often works in the mass insertion approximation
\cite{lavignac-insertion}, which allows to relate the rates for LFV processes
to the off-diagonal entries of the slepton mass matrices\footnote{The mass
insertion approximation consists in expanding the off-diagonal slepton
propagators around the diagonal, treating the off-diagonal entries of
the slepton mass matrices as small perturbations. Here the lepton basis
is the mass eigenstate basis, and the slepton basis is such that the
slepton-lepton-chargino (-neutralino) couplings are diagonal, so that all
lepton flavour violation is encapsulated in the $\delta$'s.}:
\begin{equation}
  \delta^{LL}_{\alpha \beta}\ \equiv\ \frac{(m^2_{\tilde L})_{\alpha \beta}}
  {m^2_L}\ , \quad\ \delta^{RR}_{\alpha \beta}\ \equiv\
  \frac{(m^2_{\tilde e})_{\alpha \beta}}{m^2_R}\ , \quad
  \delta^{RL}_{\alpha \beta}\ \equiv\ \frac{A^e_{\alpha \beta} v_d}{m_R m_L}\ ,
  \quad \delta^{LR}_{\alpha \beta}\ \equiv\ \delta^{RL\star}_{\beta \alpha}\, ,
  \quad\ (\alpha \neq \beta)
\label{eq:lavignac-deltas}
\end{equation}
where $m^2_L$ (resp. $m^2_R$) is the average left (resp. right) slepton
mass. For instance, the branching ratio for $\mu \rightarrow e \gamma$
is given by, at leading order \cite{lavignac-mueg-susy,lavignac-savoy02}:
\begin{equation}
  \mbox{BR}\, (\mu \rightarrow e \gamma)\ =\ \frac{3 \pi \alpha^3}
  {4 G^2_F \cos^4 \theta_W}\ \left\{ \left| f_{LL}\, \delta^{LL}_{12}
  + f_{LR}\, \delta^{LR}_{12} \right|^2 + \left| f_{RR}\, \delta^{RR}_{12}
  + f^\star_{LR}\, \delta^{LR \star}_{21} \right|^2 \right\}\ \tan^2 \beta\ ,
\label{eq:lavignac-mueg-susy}
\end{equation}
where $f_{LL}$, $f_{RR}$ and $f_{LR}$ are functions of the superpartner
masses and of the ratio of the vevs of the two MSSM Higgs doublets,
$\tan \beta =\, <\! H^0_u\! >\! /\! <\! H^0_d\! >$. For moderate and
large values of $\tan \beta$, say $\tan \beta > 10$,
$\mbox{BR}\, (\mu \rightarrow e \gamma)$ approximately scales as
$\tan^2 \beta$, unless $|\delta^{LR}_{12 (21)}| \gg
|\delta^{LL}_{12}|, |\delta^{RR}_{12}|$. From the present experimental limit
on $\mbox{BR}\, (\mu \rightarrow e \gamma)$, one can extract upper bounds on
the $\delta_{12}$'s as functions of the supersymmetric parameters. This is
illustrated in Fig. \ref{fig:lavignac-delta12}, in which mSUGRA relations
between the soft terms, i.e. universality relations at the Planck scale,
have been assumed. We can see that the  $\delta_{12}$'s are constrained to be
rather small, unless the supersymmetric partners are very heavy
\cite{lavignac-delta-bounds}. The upper bounds on the $\delta_{23}$'s and
$\delta_{13}$'s, which come from the present experimental limit on
$\mbox{BR}\, (\tau \rightarrow \mu \gamma)$ and
$\mbox{BR}\, (\tau \rightarrow e \gamma)$, respectively, are significantly
weaker, though quite stringent. This shows that supersymmetry can lead to
observable LFV processes, but their rates are controlled by the mechanism
responsible for supersymmetry breaking rather than by the parameters
associated with neutrino masses. The requirement that the corresponding
rates are below the experimental limits actually constitutes a strong
constraint on supersymmetry breaking.

\begin{figure}[htb]
\begin{center}
\epsfig{file=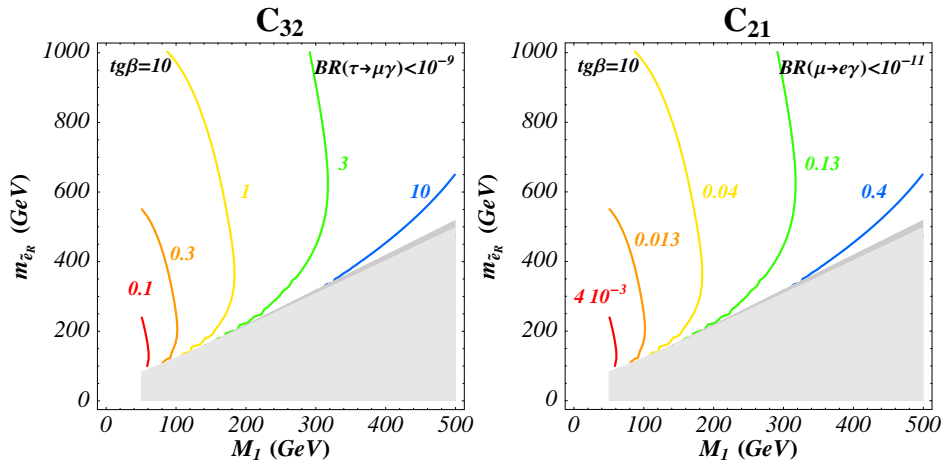,height=65mm}
\caption{Upper bounds on $|C_{23}|$ and $|C_{12}|$ associated with,
respectively, $\mbox{BR}\, (\tau \rightarrow \mu \gamma) < 10^{-9}$ and
$\mbox{BR}\, (\mu \rightarrow e \gamma) < 10^{-11}$,
as functions of the average right slepton mass $m_{\tilde e_R}$ and of the
bino mass $M_1$, for $\tan \beta = 10$ and $A_0 = m_0$. 
From Ref. \cite{lavignac-LMS01}.}
\label{fig:lavignac-Cij}
\end{center}
\end{figure}

Interestingly, the above conclusion can be evaded if the mechanism responsible
for supersymmetry breaking is flavour-blind, as might be necessary in order
to satisfy all constraints from quark and lepton flavour violating processes.
If neutrino masses are generated from the seesaw mechanism
\cite{lavignac-seesaw}, the LFV rates are then controlled by the seesaw
parameters \cite{lavignac-borzumati86}, as we discuss now.
In the seesaw mechanism, the smallness of neutrino masses naturally arises
from the couplings of the ordinary LH neutrinos to heavy Majorana RH
neutrinos:
\begin{equation}
  Y_{k \alpha}\, \bar N_{Rk} L_\alpha H_u\ +\ \frac{1}{2}\, (M_R)_{kl}\,
  N^T_{Rk} C N_{Rl}\ +\ \mbox{h.c.}\ ,
\end{equation}
where $Y$ is the Dirac mass matrix and $M_R$ the RH neutrino Majorana mass
matrix. At low energy, the effective light neutrino mass matrix is given
by (from now on, we work in the bases in which both $M_R$ and the charged
lepton mass matrix $M_e$ are diagonal):
\begin{equation}
  M_\nu\ =\ - Y^T M^{-1}_R\, Y\ =\ U^\star\, \mbox{Diag} (m_{\nu_1}, m_{\nu_2},
    m_{\nu_3})\, U^\dagger\ .
\end{equation}
For Dirac couplings of order one, $m_{\nu_3} \simeq \sqrt{\Delta m^2_{atm}}$
is obtained for right-handed neutrino masses of the order of
$5 \times 10^{14}$ GeV, remarkably close to the scale at which gauge couplings
unify in the MSSM ($2 \times 10^{16}$ GeV). Now if at some high scale
$M_U > M_R$ the soft supersymmetry breaking masses are universal in the
slepton sector:
\begin {equation}
  (m^2_{\tilde L})_{\alpha \beta} = (m^2_{\tilde e})_{\alpha \beta} =
  m^2_0\, \delta_{\alpha \beta}\ , \qquad\
  A^e_{\alpha \beta} = A_0\, Y^e_{\alpha \beta}\ ,
\end{equation}
the Dirac couplings $Y_{k \alpha}$, which violate lepton flavour, will induce
flavour off-diagonal entries through loops of heavy RH neutrinos. One thus
obtains, at low energy:
\begin {equation}
  (m^2_{\tilde L})_{\alpha \beta}\ \simeq\ -\, \frac{3 m^2_0 + A^2_0}{8 \pi^2}\
  C_{\alpha \beta}\ , \qquad (m^2_{\tilde e})_{\alpha \beta}\ \simeq\ 0\ ,
  \qquad A^e_{\alpha \beta}\ \simeq\ -\, \frac{3}{8 \pi^2}\,
  A_0 y_{e \alpha}\, C_{\alpha \beta}\ ,
\label{eq:lavignac-RGE}
\end{equation}
where the coefficients $C_{\alpha \beta} \equiv \sum_k Y^\star_{k \alpha}
Y_{k \beta} \ln (M_U / M_k)$ encapsulate the dependence on the seesaw
parameters. Putting back Eq. (\ref{eq:lavignac-RGE}) into Eqs.
(\ref{eq:lavignac-deltas}) and (\ref{eq:lavignac-mueg-susy}), one sees
that $\mbox{BR}\, (\mu \rightarrow e \gamma) \propto |C_{12}|^2$,
$\mbox{BR}\, (\tau \rightarrow \mu \gamma) \propto |C_{23}|^2$ and so on.
Fig. \ref{fig:lavignac-Cij} shows the upper bound on $|C_{12}|$
associated with the present experimental limit on
$\mbox{BR}\, (\mu \rightarrow e \gamma)$, and the upper bound on $|C_{23}|$
that would be obtained if the experimental limit on
$\mbox{BR}\, (\tau \rightarrow \mu \gamma)$ were improved by 3 orders of
magnitude.

Since the neutrino mass matrix $M_\nu$ and the coefficients $C_{\alpha \beta}$
depend on different combinations of the seesaw parameters, it is not
possible to relate the rates of LFV processes to the observed values of
neutrino masses and mixing angles. Rather experimental limits on LFV
processes can be used to discriminate between different classes of seesaw
models. In particular, in models in which the heaviest right-handed neutrino 
contributes significantly to the atmospheric neutrino mass scale, the
relation $|C_{23}| \sim |Y_{33}|^2 \ln (M_U/M_3)$ holds. Assuming that
$Y_{33}$ is of order one, as happens e.g. in $SO(10)$ Grand Unified Theories,
this implies that $\mbox{BR}\, (\tau \rightarrow \mu \gamma) > 10^{-9}$ over
a large portion of the MSSM parameter space \cite{lavignac-LMS01}. Therefore,
if the experimental sensitivity reaches $10^{-9}$ and still
$\tau \rightarrow \mu \gamma$ is not observed, this class of models will be
disfavoured over most of the MSSM parameter space. As for
$\mu \rightarrow e \gamma$, its branching ratio is generally predicted to be
large in supersymmetric seesaw models (see e.g. Refs.
\cite{lavignac-mueg-susy,lavignac-large-mueg}), with good chances of being
detected in forthcoming experiments, but it is much more model-dependent
\cite{lavignac-LMS01}.

%%%%%%%%%%%%%%%%%%%%%%%%%%%%%%%%%%%%%%%%%%%%%%%%%%%%%%%%%%%%%%%%%%%%%%%%%%%
%%%%%%%%%%%%%%%%%%%%%%%%%%%%%%%%%%%%%%%%%%%%%%%%%%%%%%%%%%%%%%%%%%%%%%%%%%%

%%%%%%%%%%%%%%%%%%%%%%%%%%%%%%%%%%%%%%%%%%%%%%%%%%%%%%%%%%%%%%%%%%%%%%%%

\section{Dipole electric moments (EDMs) of charged leptons}
%----------------------------------------------------------

While charged lepton EDMs arise only at the multiloop level in the Standard
Model, and are out of reach of foreseen experiments, they are generated at
the one-loop level in its supersymmetric extensions \cite{lavignac-EDM-susy}
and can have much larger values. One can distinguish between two types of
contributions:

\begin{itemize}
\item flavour conserving contributions from phases in flavour diagonal
parameters, i.e. the $A$-terms and the supersymmetric Higgs mass parameter
$\mu$;
\item flavour violating contributions from phases in the $\delta$'s
(off-diagonal entries of the slepton mass matrices).
\end{itemize}
At present, only the experimental limit on the electron EDM yields significant
constraints on these phases, but future experiments may be sensistive to
values of the muon EDM that are typically expected in supersymmetric models.

%%%%%%%%%%%%%%%%%%%%%%%%%%%%%%%%%%%%%%%%%%%%%%%%%%%%%%%%%%%%%%%%%%%%%%%%%%%
\begin{figure}[htb]
\begin{center}
\epsfig{file=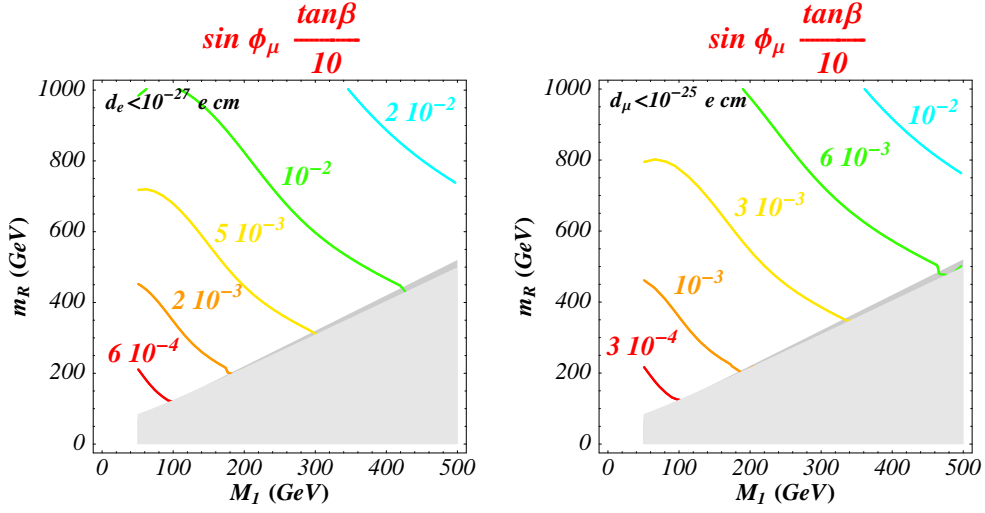,height=70mm}
\caption{Upper bounds on $\sin \Phi_\mu \frac{\tan \beta}{10}$
associated with $|d_e| < 10^{-27}$ e.cm and $|d_\mu| < 10^{-25}$ e.cm,
as functions of the average right slepton mass $m_{\tilde e_R}$
and of the bino mass $M_1$. From Ref. \cite{lavignac-savoy02}.}
\label{fig:lavignac-phi-mu}
\end{center}
\end{figure}
%%%%%%%%%%%%%%%%%%%%%%%%%%%%%%%%%%%%%%%%%%%%%%%%%%%%%%%%%%%%%%%%%%%%%%%%%%%

The flavour conserving contributions are proportional either to
$\mbox{Im} (A_i - \mu \tan \beta)\, m_{e_i}$ or to
$\mbox{Im} (\mu) \tan \beta\, m_{e_i}$.
Unless there is a strong hierarchy among the $A_i$, it follows that the
charged lepton EDMs approximately satisfy the scaling relation
$d_i \propto m_{e_i}$, where $d_i$ is the EDM of the $i^{\rm th}$ charged
lepton, and $m_{e_i}$ is its mass. The experimental limit on the electron
EDM, $|d_e| < 1.6 \times 10^{-27}$ (90 \% C.L.) \cite{lavignac-regan02},
strongly constrains the phase of the $\mu$ parameter (see Fig.
\ref{fig:lavignac-phi-mu}), while the constraint on $\mbox{Im} A_e$ is
much weaker. Note that the upper bound on $\sin \Phi_\mu$ cannot be
relaxed by lowering the value of $|\mu|$, since the latter is constrained
by the condition of electroweak symmetry breaking.
The present experimental limits on $d_\mu$ and $d_\tau$ do not yield any
significant constraint on $\sin \Phi_\mu$ and $\mbox{Im} A_{\mu, \tau}$.
Moreover, the scaling relation $d_i \propto m_{e_i}$, together with the
experimental limit on the electron EDM, implies a strong upper bound
on the muon EDM, which is smaller by 7 orders of magnitude than the present
experimental limit and lies below the sensitivity of the planned BNL
experiment \cite{lavignac-BNL-dmu}:
\begin{equation}
 \left. d_\mu \right|_{\rm th (FC)}\ <\ \frac{m_\mu}{m_e}\ \left.
  d_e \right|_{\rm exp}\ =\ 3 \times 10^{-25}\, \mbox{e.cm}\ .
\end{equation}
This fact is illustrated in Fig. \ref{fig:lavignac-phi-mu}, where the
upper bound on $\sin \Phi_\mu$ that would correspond to a limit of
$10^{-25}$ e.cm on the muon EDM (right) is compared with the upper bound
obtained from the present experimental limit on the electron EDM (left).

Since EDMs are flavour diagonal quantities, the flavour violating
contributions necessarily involve two $\delta$'s. For instance, the
following combinations contribute to $d_i$:
$\sum_k \mbox{Im}\, (\delta^{LL}_{ik} \delta^{LR}_{ki})$,
$\sum_k \mbox{Im}\, (\delta^{LR}_{ik} \delta^{RR}_{ki})$,
$\sum_k m_k\, \mbox{Im}\, (\delta^{LL}_{ik} (A^\star_k - \mu \tan \beta)
\delta^{RR}_{ki})$,
$\sum_k m_k\, \mbox{Im}\, (\delta^{LR}_{ik} (A_k - \mu^\star \tan \beta)
\delta^{LR}_{ki})$.
These contributions do not satisfy the scaling relation
$d_i / d_j \approx m_{e_i} / m_{e_j}$; in particular, it is possible
to have $d_\mu \gg \frac{m_\mu}{m_e} \left. d_e \right|_{\rm exp}$, in the
sensitivity range of the future BNL experiment \cite{lavignac-large-dmu}.
Note that the experimental upper limit on the electron EDM provides better
constraints on the (imaginary part of the) products $\delta_{13} \delta_{31}$
than the LFV decay $\tau \rightarrow e \gamma$.

Let us finally add that, in supersymmetric seesaw models with flavour-blind
supersymmetry breaking, the phases present in the Dirac couplings can induce
complex off-diagonal slepton masses, connecting the values of the charged
lepton EDMs to the seesaw parameters.

%%%%%%%%%%%%%%%%%%%%%%%%%%%%%%%%%%%%%%%%%%%%%%%%%%%%%%%%%%%%%%%%%%%%%%%%

\section{Conclusions}
%--------------------

We have seen in this short review that flavour and CP violation have a
very different status in the lepton and in the quark sectors. 
Indeed, if the PMNS matrix is the only source of flavour and CP violation
in the lepton sector, neutrino oscillations are likely to be the only
manifestation of lepton flavour violation that can be accessed experimentally,
as well as the only place (together with neutrinoless double beta decay if
neutrinos are Majorana particles) where one can possibly 
test leptonic CP violation.

LFV processes involving charged leptons and charged lepton EDMs are
therefore a unique probe of new physics. The observation of e.g.
$\mu \rightarrow e \gamma$, or the
measurement of a nonzero muon EDM would definitely testify
for physics beyond the Standard Model. A significant improvement of 
the experimental upper limits on $\mbox{BR}\, (\mu \rightarrow e \gamma)$
and $\mbox{BR}\, (\tau \rightarrow \mu \gamma)$ would already provide strong
constraints on supersymmetry breaking and on supersymmetric seesaw models.

%%%%%%%%%%%%%%%%%%%%%%%%%%%%%%%%%%%%%%%%%%%%%%%%%%%%%%%%%%%%%%%%%%%%%%%%

%%%%%%%%%%%%%%%%%%%%%%%%%%%%%%%%%%%%%%%%%%%%%%%%%%%%%%%%%%%%%%%%%%%%%%%%

\label{LavignacEnd}
 
\end{document}